\documentclass[12pt,preprint]{aastex}
\input epsf.sty

\begin{document}
\title{Hydromagnetic instabilities in proto--neutron stars} 
\author{Juan A. Miralles$^{1)}$, Jos\'e A. Pons$^{2)}$ and Vadim A. Urpin$^{3,4)}$} 
\affil{$^{1)}$ Departament de F\'{\i}sica Aplicada, Universitat d'Alacant, 
                    Ap. Correus 99, E-03080 Alacant, Spain \\ 
           $^{2)}$ Dipartimento di Fisica ``G. Marconi'',  
		   Universita' di Roma ``La Sapienza'', I-00184,
                    Roma, Italy \\ 
           $^{3)}$ Departament d'Astronomia i Astrof\'{\i}sica,  
		   Universitat de Val\`encia, 
                    E-46100 Burjassot, Spain \\ 
           $^{4)}$ A.F.Ioffe Institute of Physics and Technology, 
                    194021 St Petersburg, Russia} 
 
\date{Received.....; accepted.....} 
\markboth{J.A.Miralles, J.A.Pons \& V.Urpin: Hydromagnetic instabilities in 
           proto--neutron stars}{}

\begin{abstract} 
 
The stability properties of newly born neutron stars, or 
proto--neutron stars (PNSs), are considered. We take
into account dissipative processes, such as neutrino transport and viscosity,
in the presence of a magnetic field. 
In order to find the regions of the star subject to different 
sorts of instability, we derive the general instability 
criteria and apply it to evolutionary models of PNSs. 
The influence of the magnetic field on instabilities is analyzed 
and the critical magnetic field stabilizing the star is obtained. 
In the light of our results, we estimate of the maximum 
poloidal magnetic field that might be present in 
young pulsars or magnetars.
\end{abstract} 


\section{Introduction}  

Neutron stars are born in the aftermath of core collapse Supernovae
explosions, the fate of massive stars within a range of 8 to 30$M_{\odot}$.
In the final stages of stellar evolution an iron core grows, until
electron captures or photodesintegration of heavy nuclei trigger its 
gravitational collapse. The collapse
proceeds until nuclear density is reached and the stiffening of the
equation of state (EOS) provokes the bounce of the infalling material.
A shock wave forms and propagates outwards, leaving behind a newly born,
hot, lepton--rich PNS. These objects are prodigious
emitters of neutrinos of all types, which dominate the energetics of core
collapse Supernovae and play a crucial role in the explosion mechanism
by carrying away the binding energy from the PNS and depositing a part 
of it in the outer regions.

Whether or not the energy relocated by neutrinos is able to relaunch
the stalled shock is still a controversial issue
\citep{Bur00,RJ00,Jan00,Lie01}
that has led to speculation about alternative ways the explosion is powered
such as convection, rotation, and magneto-hydrodynamics effects.
In particular, it has been recognized by a number of authors that
convective energy transport in the newly born PNS might play an
important role by increasing the neutrino luminosity required to
ignite the Supernova explosion 
\citep{BF93,JM96,BD96}
Convection in PNSs can be driven not only by the lepton gradient,
as originally suggested by \cite{Eps79}, but also by the development
of negative entropy gradients, which is common in many simulations
of supernovae models \citep{BM94,BMD95,RJ00,Lie01}
and evolutionary models of PNSs \citep{BL86,KJ95,KJM96,Pons99,Pons01a,Pons01b}
despite differences in the equation of state and neutrino transport. 

Although stellar core collapse may also have dramatic consequences 
on the magnetic field strength, the effect of magnetic fields was 
not considered in the above mentioned works,
mainly due to the tough technical difficulties inherent to multidimensional
magnetohydrodynamical simulations.
It is known that the magnetic field in a collapsing star can be 
amplified by many orders of magnitude due to the conservation of the magnetic  
flux. This idea is often used to explain the origin of strong magnetic  
fields in neutron stars \citep{Gin64,GO64,Wol64,ZN71,ST83}. The conductivity
of plasma inside the star is so large that the magnetic decay timescale 
is much shorter than the collapse time for a lengthscale of the magnetic
field comparable to the stellar radius. Therefore, the field is frozen-in, 
and the magnetic flux is conserved during the collapse stage. In the case 
of a frozen-in magnetic field, the average field strength increases 
approximately as $B \propto \rho^{2/3}$ where $\rho$ is the density. 
By assuming that the magnetic field strength of the core of a massive star is
comparable to that of strongly magnetized  white dwarfs ($\sim 10^{9}$ G), at 
the end of core collapse, the neutron star may have a magnetic field 
$\sim 10^{13} - 10^{14}$ G, comparable to that observed in young radio-pulsars.
Following the same arguments, 
if we consider that the magnetic flux is conserved during the evolution of a 
massive star from the main sequence, starting from 
an initial magnetic field of $\sim 10^{3}-10^{4}$ G, we 
end up with a magnetic field of $\sim 10^{13}-10^{14}$ G at the time the 
core of the star has become a neutron star.
Similar arguments are often used to estimate the magnetic field of neutron 
stars formed from magnetized white dwarfs by the {\it accretion induced 
collapse} mechanism.
Note that these estimates give the order of magnitude of  
a poloidal field. The toroidal field, however, might be  much stronger  
\citep{ABP80}.

Under certain conditions, convective motions in PNSs can amplify the
magnetic field via dynamo action. The most optimistic estimates
lead some authors \citep{TD93,TM01} to the conclusion
that turbulent dynamo action could be responsible for the formation of 
ultra-magnetized neutron stars (magnetars), with fields as strong as 
$\sim 10^{16}$ G at a very early evolutionary stage.
Certainly, such strong magnetic fields influence magneto-hydrodynamic 
processes in PNSs, particularly their stability properties.  
At this point, the fact that the convective overturn is intrinsically 
a 3D phenomenon makes fully consistent numerical simulations 
a very laborious and computationally expensive way to address the problem,
and simplified semi-analytical approaches 
are extremely useful to evaluate the necessity of more detailed 
studies. In this respect, some attempts have been made to establish definite 
criteria as
to the different types of instability and their growth times \citep{BD96,MPU00}.
In MPU we derived the general criteria for instability, taking dissipative 
processes into account such as neutrino transport and 
viscosity. In the present paper, we consider the stability  
properties of magnetic PNSs in detail. It is well known (see,   
e.g., Chandrasekhar 1961) that the magnetic field may stabilize a fluid   
against convection causing an additional effective viscosity due to   
Lorentz force. On the other hand, the field can lead to new branches of   
instability associated, for instance, with the Alfv\'en waves. In this   
paper, we present the main results of a general stability analysis in   
magnetized PNSs. The stability properties will be considered for a wide 
range of magnetic fields.  
 
The paper is organized as follows. In \S 2, 
we obtain the dispersion equation which determines the
condition of linear instability from the equations 
governing magneto-hydrodynamics of PNSs and neutrino transport.
In \S 3, we discuss the stability properties of both non-magnetic 
(\S 3.1) and magnetic (\S3.2) PNSs,
appling the criteria of hydromagnetic stability 
to the results of numerical simulations to identify 
the unstable zones. We also calculate the critical magnetic field that 
stabilizes a PNS. Finally, in \S 4 we summarize and discuss
our main findings.
 
\section{The dispersion equation}  
 
Consider the stability properties on adopting a plane-parallel geometry.  
We assume that within the layer between $z=0$ and $z=d$ the gravity   
${\bf g}$ is directed in the negative $z$-direction and neglect a   
non-uniformity of ${\bf g}$ as well as general relativistic corrections to the 
equations of hydrodynamics. The characteristic cooling time scale of a   
PNS is assumed to be much longer than the growth time   
of instability thus the latter can be treated in a quasi-stationary   
approximation. Since the convective velocities are typically much   
smaller than the speed of sound, one can describe instability by making   
use of the standard Boussinesq approximation (see, e.g., Landau   
\& Lifshitz 1987). We consider the linear instability regime when the 
equations   
governing small perturbations can be obtained by the linearization of   
hydrodynamic equations. In the following, small perturbations of hydrodynamic   
quantities will be marked by a subscript ``1''. For simplicity,  
the unperturbed magnetic field is assumed to be vertical.  
The linearized momentum, continuity and induction equations read  
\begin{equation}  
\rho \dot{{\bf v}}_{1} = - \nabla p_{1} + {\bf g} \rho_{1}+ \frac{1}{4 \pi}  
(\nabla \times {\bf B}_{1}) \times {\bf B} + \rho \nu \Delta {\bf v}_{1},  
\end{equation}  
\begin{equation}  
\nabla \cdot  {\bf v}_{1} = 0,  
\end{equation}  
\begin{equation}  
\dot{{\bf B}}_{1} = \nabla \times ({\bf v}_{1} \times {\bf B}) + \nu_{m}  
\Delta {\bf B}_{1},  
\end{equation}  
\begin{equation}  
\nabla \cdot {\bf B}_{1} = 0,  
\end{equation}  
where $p$ and $\rho$ are the pressure and density, respectively, $\nu$ is  
the viscosity and $\nu_{m}$ is the magnetic diffusivity. We assume that the 
matter inside a PNS is in chemical equilibrium, thus the 
density can generally be considered as a function of the pressure $p$, 
temperature $T$ and lepton fraction $Y$, $Y=(n_{e} + n_{\nu})/n$, with $n_e$
and $n_\nu$ being the net (particles minus antiparticles) number densities
of electrons and neutrinos, respectively, and $n=n_{p}+n_{n}$ is the 
number density of baryons. Then, the perturbations of density, entropy 
per baryon ($s$) and neutrino chemical potential ($\mu$) can be expressed 
in terms of $p_{1}$, $T_{1}$ and $Y_{1}$. In the Boussinesq approximation, 
the perturbations of pressure are negligible because the fluid motions 
are assumed to be slow and the moving fluid elements are nearly in pressure 
equilibrium with their surroundings. Therefore, we have  
\begin{equation}  
\rho_{1} \approx - \rho \left( \beta \frac{T_{1}}{T} + \delta Y_{1} \right),  
\end{equation}  
\begin{equation}  
s_{1} \approx m_{p} c_{p} \frac{T_{1}}{T} + \sigma Y_{1},  
\end{equation}  
where $m_{p}$ is the proton mass (we neglect the mass difference between   
protons and neutrons), $\beta$ and $\delta$ are the coefficient of   
thermal and chemical expansion, respectively, $\beta = - (\partial   
\ln \rho/\partial \ln T)_{pY}$, $\delta = - (\partial   
\ln \rho/ \partial Y)_{pT}$; $\sigma = (\partial s /\partial Y)_{pT}$;  
and $c_{p} = (T/m_{p}) (\partial s / \partial T)_{pY}$ is the specific   
heat at constant pressure.
 
The above equations should be complemented by the equation driving  
the evolution of chemical composition and heat balance. We employ the   
equilibrium diffusion approximation, which is sufficiently accurate
and reliable during the early stage of PNS evolution,  when the   
mean free path of neutrino is short compared to the density and  
temperature length scales. In this approximation, the diffusion and   
thermal balance equations can be linearized (see MPU for details)
and written as
\begin{equation}  
\dot{Y}_{1} + {\bf v}_{1} \cdot \nabla Y =  \lambda_{T}   
\frac{\Delta T_{1}}{T} + \lambda_{Y} \Delta Y_{1},  
\label{lint1}  
\end{equation}  
\begin{equation}  
\frac{\dot{T}_{1}}{T} - {\bf v}_{1} \cdot \frac{\Delta \nabla T}{T} =   
\kappa_{T} \frac{\Delta T_{1}}{T} +  
\kappa_{Y} \Delta Y_{1},  
\label{lint2}  
\end{equation}  
where  
\begin{equation}  
\Delta \nabla T \equiv - \frac{T}{m_{p} c_{p}} ( \nabla s -   
\sigma \nabla Y ) = \left( \frac{\partial T}{\partial p} \right)_{s,Y}  
\nabla p - \nabla T  
\end{equation}  
is the superadiabatic temperature gradient, i.e., the difference  
between the temperature gradient of a fluid with constant entropy and  
composition and the actual temperature gradient, and 
the transport coefficients $\lambda_T, \kappa_T, \lambda_Y, \kappa_Y$
have been defined in MPU.

Equations (1)--(4), (\ref{lint1}) and  (\ref{lint2})
together with the corresponding   
boundary conditions determine the behaviour of small   
perturbations. For simplicity, we consider the case of vanishing 
perturbations at the boundaries $z=0$ and $z=d$. Note that  
other boundary conditions cannot change the main   
conclusions of our analysis qualitatively.  
    
Without loss of generality,  
the dependence of all perturbations on time and on the horizontal  
coordinate can be chosen as $\exp(\gamma t - i k x)$, 
where $k$ is the horizontal wavevector and $\gamma$ is a constant 
which can be complex and its real part gives the inverse growth 
(or decay) timescale. Then,
equations (1)-(4), (12), and (13) can be reduced to  
only one equation of higher order. Solving for $v_{1z}$, we obtain  
\begin{eqnarray}  
\Delta [(\gamma - \kappa_{T} \Delta )(\gamma - \lambda_{Y} \Delta) -  
\lambda_{T} \kappa_{Y} \Delta^{2}]  
\nonumber \\	 
\left[ (\gamma - \nu \Delta)  (\gamma - \nu_{m} \Delta)  
- c_{A}^{2} \frac{d^{2}}{dz^{2}} \right]   
v_{1z} =  
\nonumber \\ 
gk^{2} \left\{   
\frac{d Y}{dz} [\delta (\gamma - \kappa_{T} \Delta) + \beta \kappa_{Y} \Delta ] 
\right. \nonumber \\  
\left.  - \frac{\Delta \nabla T}{T}   
[\beta (\gamma - \lambda_{Y} \Delta) + \delta \lambda_{T} \Delta]   
\right\} (\gamma - \nu_{m} \Delta) v_{1z}, 
\end{eqnarray}  
where $c_{A} = B / \sqrt{4 \pi \rho}$ is the Alfv\'en velocity and
$\Delta = \frac{d^{2}}{d z^{2}} - k^{2}.$  
The coefficients of this equation are constant in our simplified approach.  
The solution for the fundamental mode with ``zero boundary  
conditions'' has the form $v_{1z} \propto \sin (\pi z/ d)$. Thus 
the dispersion equation for the fundamental mode is  
\begin{eqnarray}  
[ (\gamma + \omega_{\nu}) (\gamma + \omega_{m}) + \omega_{A}^{2}]   
[(\gamma + \omega_{T})  
(\gamma + \omega_{Y}) - \omega_{TY} \omega_{YT}] &=&  
\nonumber \\  
(\gamma + \omega_{m}) \left[\omega_{g}^{2} ( \gamma + \omega_{Y} - \alpha   
\omega_{YT} ) +   
\omega_{L}^{2} \left( \gamma + \omega_{T} -   
\frac{\omega_{TY}}{\alpha} \right) \right], &&  
\label{dis1}
\end{eqnarray}  
where $\alpha = \delta / \beta$. In this expression, we introduced   
the characteristic frequencies  
\begin{eqnarray}  
&&\omega_{\nu}= \nu Q^{2}, \omega_{m}= \nu_{m} Q^{2}, \; \omega_{T} =   
\kappa_{T} Q^{2}, \; \omega_{Y} = \lambda_{Y} Q^{2},     
\nonumber \\  
&&\omega_{YT} = \lambda_{T} Q^{2}, \;  
\omega_{TY} = \kappa_{Y} Q^{2}, \omega_{A} = c_{A} \pi/a , \\  
&&\omega_{g}^{2} = \frac{\beta g k^{2} \Delta \nabla T}{T Q^{2}}, \;   
\omega_{L}^{2} = - \frac{\delta g k^{2}}{Q^{2}} \cdot   
\frac{d Y}{d z},        \nonumber  
\end{eqnarray}  
with $Q^{2} = (\pi/d)^{2} + k^{2}$. The quantities $\omega_{\nu}$,   
$\omega_{m}$, $\omega_{T}$, and $\omega_{Y}$ are the inverse time   
scales of dissipation of the perturbations due to viscosity, magnetic  
diffusivity, thermal conductivity and chemical diffusivity, respectively;   
$\omega_{YT}$ characterizes the rate of diffusion caused by the temperature   
inhomogeneity (thermodiffusion), and $\omega_{TY}$ describes the influence   
of chemical inhomogeneities on the rate of thermal evolution;  
$\omega_{g}$ is the frequency (or, in the case of instability, the  
inverse growth time) of the buoyant wave; $\omega_{L}$ characterizes   
the dynamical time scale of the processes associated with the lepton   
gradient, and $\omega_{A}$ is the Alfv\'en frequency.  

\section{Stability of proto-neutron stars}
  
\subsection{Non-magnetic PNSs}

Consider initially the case when   
${\bf B}={\bf 0}$. Then, equation (\ref{dis1}) is reduced to a cubic equation   
with the roots given by   
\begin{equation}  
\gamma^{3} + a_{2} \gamma^{2} + a_{1} \gamma + a_{0} = 0,  
\label{cubic}  
\end{equation}  
where   
\begin{eqnarray}  
a_{2} &=& \omega_{\nu} + \omega_{T} + \omega_{Y},  \nonumber  \\  
a_{1} &=& \omega_{T} \omega_{Y} - \omega_{TY} \omega_{YT} +  
\omega_{\nu} (\omega_{T} + \omega_{Y}) - (\omega_{g}^{2} +  
\omega_{L}^{2}),  \nonumber  \\  
a_{0} &=& \omega_{\nu} (\omega_{T} \omega_{Y} - \omega_{TY} \omega_{YT})  
\nonumber  \\  
&-& \omega_{g}^{2} ( \omega_{Y} - \alpha \omega_{YT})   
- \omega_{L}^{2} \left( \omega_{T} - \frac{1}{\alpha} \omega_{TY} \right).  
\end{eqnarray}  
  
Equation (\ref{cubic}) describes three essentially different modes,
corresponding to the three roots of the equation, 
which can generally exist in a chemically inhomogeneous fluid. The   
condition that at least one of the roots has a positive real part   
(unstable mode) is equivalent to one of the following inequalities  
\begin{equation}  
a_{2} < 0, \;\;\; a_{0} < 0, \;\;\; a_{1} a_{2} < a_{0}  
\label{3cond}  
\end{equation}   
being fulfilled (see, e.g., Aleksandrov, Kolmogorov, \& Laurentiev 1985; 
DiStefano III, Stubberud, \& Williams 1994).  
Since $\kappa_T$ and $\lambda_Y$ are positive defined quantities,   
the first condition $a_2<0$ will never apply, and only the two other  
conditions determine the stability of non-magnetic PNSs.  
  
Generally, the last two conditions (\ref{3cond}) are rather 
complex and  depend on the horizontal wavevector of perturbations, $k$. 
The critical temperature and lepton gradients which determine the 
transition from stability to instability can be quite different for 
perturbations with different $k$. In order to obtain the true criteria 
we have to find the minimal values of these gradients which can cause 
instability. From the properties of the kinetic coefficients we have 
$(\kappa_{T} \lambda_{Y} - \kappa_{Y} \lambda_{T}) > 0$, therefore the 
gradients in the last two conditions (\ref{3cond}) are minimal for 
perturbations with the wavevector $k$ minimizing the quantity 
$Q^{6}/k^{2}$. This happens when $k^{2} = (\pi/d)^{2}/2$.
For such a horizontal wavevector, the instability criteria read  
\begin{eqnarray}  
\frac{dY}{dz} (\delta \kappa_{T} &-& \beta \kappa_{Y} ) -  
\frac{\Delta \nabla T}{T} (\beta \lambda_{Y} - \delta \lambda_{T})    
\nonumber \\
&<& - \frac{27 \pi^{4}}{4 g d^{4}} \nu (\kappa_{T}   
\lambda_{Y} - \kappa_{Y} \lambda_{T}),  
\label{nfcond}  
\\  
\frac{d Y}{dz} [ \delta ( \nu + \lambda_{Y} ) +  \beta \kappa_{Y}]   
&-&  \frac{\Delta \nabla T}{T} [ \beta ( \nu + \kappa_{T} ) +   
\delta \lambda_{T}]  
\nonumber \\
&<& - \frac{27 \pi^{4}}{4 g d^{4}} (\kappa_{T} + \lambda_{Y})  
[ (\nu + \kappa_{T}) (\nu + \lambda_{Y}) - \lambda_{T} \kappa_{Y}] .  
\label{smcond}  
\end{eqnarray}  
  
Conditions (\ref{nfcond})--(\ref{smcond}) divide the $\Delta 
\nabla T - \nabla Y$ plane into four regions, which are characterized 
by different stability properties. The size and configuration of these 
regions depend on the thermodynamic and kinetic properties of nuclear
matter. The detailed analysis of stability of non-magnetic PNSs based 
on criteria (\ref{nfcond}) and (\ref{smcond}) has been  done in MPU. 
In that paper, as in the present one, we use results from numerical 
simulations of PNS evolution by \cite{Pons99} to calculate   
the different thermodynamical derivatives, diffusion coefficients,   
and conductivities appearing in the stability criteria. To illustrate 
the large variety of situations encountered during the Kelvin-Helmholtz 
phase of a PNS, we have chosen three different cases.
The physical conditions and the different transport parameters are 
summarized in Table 1. Case A corresponds to a layer near the center 
of the star, with an enclosed mass of 0.05 $M_\odot$,  at
an evolutionary time of $t=2$ s, when the PNS is in the deleptonization 
stage and the core is being heated. Cases B and C correspond to two 
different layers at a time of $t=20$ s, during the cooling stage, 
mainly driven by thermal neutrinos since an important part of the
lepton content of the star has already been radiated away. 
The layer corresponding to case B is located at the same depth as A
(0.05 $M_\odot$) star while the layer in case C has an enclosed 
baryonic mass of 1 $M_\odot$.  In our calculations 
we have used the value of $d$ given by the pressure height 
$d=\ell \equiv \mid d\ln p/dr\mid ^{-1}$. 
  
In Figure 1 we plot the lines of critical stability given by equations    
(\ref{nfcond}) and (\ref{smcond}) (solid and dotted line respectively) 
in a $\Delta \nabla T  - \nabla Y$ plane. The asterisks indicate the 
particular values of the gradients for each of the three cases considered, 
the parameters of which are summarized in Table 1. The shaded area covers 
the region where all the roots of the cubic equation (\ref{cubic}) are real.   
We denote by {\it neutron fingers} a region where condition (\ref{nfcond}) 
is fulfilled but not condition (\ref{smcond}); by {\it convection} we 
denote those regions where both conditions (\ref{nfcond}) and (\ref{smcond}) 
are satisfied; we denote by {\it semiconvection} the case when condition 
(\ref{smcond}) is satisfied but condition (\ref{nfcond}) is not.  
  
By applying Routh criterium (DiStefano III, Stubberud, \& Williams 1994), 
which states that the number of unstable modes of a cubic equation is given  
by the number of changes of sign in the sequence   
\begin{equation}  
\left\{1,a_2,\frac{a_2a_1-a_0}{a_2},a_0\right\},  
\end{equation}  
we deduce that the number of unstable modes in the regions labelled by 
{\it convection} or {\it neutron finger} is always one. Thus, the 
root corresponding to this mode is real and positive. Notice that, 
while in the
{\it convection} case all roots are real, in the {\it neutron finger} 
case, the two stable modes can be either real or complex conjugates.
In the region labelled {\it semiconvection} we have two unstable modes, 
which may be either real or complex conjugates. {\it Semiconvection} can be, 
therefore, oscillatory (complex conjugate unstable modes) or 
non-oscillatory (two real unstable modes). We emphasize that the convention 
we use to divide the plane $\Delta \nabla T  - \nabla Y$ according to 
the instability criteria (\ref{nfcond}) and (\ref{smcond}) differs from 
that of \cite{BD96}. According to their 
convention, the convectively unstable region is extended to 
contain that part of our neutron finger and semiconvection 
regions where all roots are real. In general, the classification in 
different types of instability is somehow arbitrary, because
there is no clear qualitative distinction between {\it neutron fingers} and 
{\it convection}: in both cases the unstable root is real.
Our classification simply responds to the fact that, in the limit of 
vanishing diffusivity, criteria (\ref{nfcond}) and (\ref{smcond}) 
reduce to the classical Rayleigh-Taylor and Schwarzschild criteria, and
certainly is not the only valid choice. The important idea to bear in
mind is that, when all the kinetic coefficients are of similar magnitude, 
the distinction among the different types of instabilities is less 
evident and only the concept of overall stability is meaningful. 
  

\subsection{Magnetic PNSs}

In the general case, ${\bf B} \neq 0$, equation (18) can be rewritten as
\begin{equation}
\gamma^{4} + q_{3} \gamma^{3} + q_{2} \gamma^{2} + q_{1} \gamma 
+ q_{0} = 0,
\label{quartic}
\end{equation}
where 
\begin{eqnarray}
q_{3} &=& a_2 + \omega_{m},   \nonumber  \\
q_{2} &=& a_1 + \omega_{m} a_2 + \omega_{A}^{2},  \nonumber  \\
q_{1} &=& a_0+ \omega_{m} a_1 + \omega_{A}^{2} (\omega_{T} + \omega_{Y})  
\nonumber \\ 
q_{0} &=& \omega_{m} a_0 + \omega_{A}^{2} (\omega_{T} \omega_{Y} - 
\omega_{TY} \omega_{YT}). 
\label{qs}
\end{eqnarray}
Equation (\ref{quartic}) describes four essentially different modes 
which can generally exist in chemically inhomogeneous fluid in the 
presence of the magnetic field. The Routh criterium applied to a quartic 
equation states that the number of roots with positive real part (unstable 
modes) is given by the number of changes of sign in the sequence 
\begin{equation}
\left\{ 1,~q_3, ~\frac{q_3q_2-q_1}{q_3}, 
~q_1-\frac{q_3^2q_0}{q_3q_2-q_1}, ~q_0 \right\}.
\end{equation}
Since $q_3$ is always positive, there is, at least, one stable mode. 
The criteria for hydro--magnetic instability is equivalent to requiring
that any one of the following inequalities be satisfied:
\begin{eqnarray}
q_{0} &<& 0~, 
\label{mcr1} \\
q_3q_2-q_1 &<& 0~, 
\label{mcr2} \\ 
q_1- \frac{q_3^2q_0}{(q_3q_2-q_1)} &<& 0 ~.
\label{mcr3}
\end{eqnarray}
In general, the magnetic field has a stabilizing effect on all three 
conditions.
Consequently, there will be a critical value of ${\bf B}$ above which none 
of the instability conditions is fulfilled and all modes become stable.

To warm up, we begin analyzing the behaviour of the roots with increasing
magnetic field.  In Figure 2 we draw a sketch, plotting the real part ($\gamma$)
of the four modes as a function of $\omega_A^2$. Modes with positive $\gamma$
are unstable.  In order to show more clearly the qualitative effect,  
we have assigned values to the dissipative frequencies
which do not correspond to real values encountered in the early
stages of a neutron star's life. The qualitative behaviour using real
coefficients is very similar, although the different scales would not allow one
to capture the trends in a simple plot.
In the limit of a vanishing magnetic field ($\omega_A \rightarrow 0$)
there are three non-magnetic modes (solid lines), discussed in
the previous section, which remain essentially unaltered, and a stable
magnetic mode with a value equal to $-\omega_m$ (dashed line).
The situation shown in Figure 2 would correspond to the {\it convection}
case (1 unstable and 2 stable non-magnetic modes when  ${\bf B}=0$).
By switching on the magnetic field, the real part of the new magnetic mode 
(dashes) increases, and becomes positive in the point labelled $a$,
which corresponds to the physical conditions such that $q_0=0$.
At that point, the original unstable mode (upper solid line) has been 
partially damped.
Further increasing the magnetic field, the real parts of the 
originally unstable mode and the new unstable magnetic mode merge, to
form a pair of complex conjugate modes (point $b$). Convection becomes, thus,
oscillatory, due to the stabilizing action of the magnetic field.
At a critical magnetic field, given by condition (\ref{mcr3}),
the real part of the unstable modes 
vanishes (point $c$), and for higher magnetic field all modes are stable.
The originally stable modes are not seriously affected.

The general trends discussed above (namely: damping of the unstable mode,
a new magnetic mode becoming unstable, merging of modes and change
to oscillatory convection, and final suppression of instabilities)
are also found when realistic kinetic coefficients from numerical simulations
are used.  To quantify the characteristic values of the magnetic field 
causing the effects described above, we have taken the parameters 
corresponding to case B from Table 1. Those modes with a typical wavelength 
of the order of the pressure height (see below for further discussion), suffer 
the transitions $a$, $b$, $c$ in Figure 2 when
$B = 2.5\times 10^{11}$ G, $1.8 \times 10^{16}$ G and $2.09\times 10^{16}$ G, 
respectively. Thus, a field larger than $2.09\times 10^{16}$ G 
entirely suppresses all types of instability in the considered region.

In order to understand the effect of the magnetic field on each instability 
condition, it is more convenient to rewrite equations (\ref{mcr1}--\ref{mcr3})
separating the terms containing 
the Alfv\'en frequency and magnetic dissipation:
\begin{equation}
a_0 < - \frac{\omega_A^2}{\omega_m} (\omega_T \omega_Y -
\omega_{TY}\omega_{YT}),
\label{mc1}
\end{equation}
\begin{equation}
a_2 a_1- a_0 < -\omega_m a_2 (a_2+\omega_m) - \omega_A^2 
(\omega_\nu+\omega_m),
\label{mc2}
\end{equation}
\begin{eqnarray}
[a_0 + \omega_m a_1 + \omega_A^2 (\omega_T+\omega_Y)] 
[a_2 a_1 - a_0 +
\omega_m a_2 (a_2+\omega_m)  
+ \omega_A^2 (\omega_\nu+\omega_m)] 
\nonumber \\
< (a_2+\omega_m)^2 
\times [(\omega_m a_0+ \omega_A^2 (\omega_T \omega_Y - 
\omega_{TY}\omega_{YT})].
\label{mc3}
\end{eqnarray} 
These inequalities can be further simplified if we take 
into account that the frequencies entering these conditions are of 
different orders of magnitude in PNSs. The dissipative frequencies,
$\omega_{\nu}$, 
$\omega_{T}$, $\omega_{Y}$, $\omega_{TY}$, and $\omega_{YT}$ are 
approximately comparable to each other and range between $10-10^{3}$ s$^{-1}$.
The magnetic dissipative frequency, $\omega_{m}$, is much smaller. 
The electrical resistivity of hot nuclear matter ($\cal{R}$) is relatively 
low, ${\cal R} \approx 6 \times 10^{-29} T_{8}^{2}$ s \citep{YS91},
thus the characteristic time scale of dissipation of 
the magnetic field is extremely long compared to other dissipative 
time scales. Taking into account that $\nu_{m} = c^{2} {\cal R}/4 \pi$, we 
can estimate $\omega_{m} \sim 10^{-11}$ s$^{-1}$ for $k \sim \pi/\ell$. 
Therefore, we can neglect $\omega_{m}$ when it enters conditions 
(\ref{mcr2}) and (\ref{mcr3}) in a row with other dissipative frequencies. 
In addition, due to the small magnetic diffusivity, it also turns out that 
$\omega_m \omega_\nu \ll \omega_A^2$ and equation (\ref{mcr1}) can be written as
\begin{eqnarray}
-\omega_{g}^{2} (\omega_{Y} - \alpha \omega_{YT}) -
\omega_{L}^{2} \left( \omega_{T} - \frac{1}{\alpha} \omega_{TY}
\right)  
+ \frac{\omega_{A}^{2}}{\omega_{m}} 
(\omega_{T} \omega_{Y} - \omega_{TY} \omega_{YT} )< 0.          
\label{cq0}
\end{eqnarray}  
All three conditions depend on the wavevector $k$. It is well known 
\citep{Cha61} that in the inviscid limit the 
minimal values of gradients which can cause instability of a 
magnetized fluid correspond to $k \rightarrow \infty$. 
Taking account of viscosity increases the most unstable horizontal 
wavelength and the critical gradients are reached for a large, though
finite, value of $k$. In the case of low magnetic 
dissipation, $k$ is much larger than $\pi /d$. 
Substituting $k \approx Q$ into equation (35), we have  
\begin{eqnarray}
\frac{dY}{dz} (\delta \kappa_{T} - \beta \kappa_{Y} ) -
\frac{\Delta \nabla T}{T} (\beta \lambda_{Y} - \delta \lambda_{T})  
< - \frac{\pi B^{2}}{4 g \rho d^{2} \nu_{m}} (\kappa_{T} 
\lambda_{Y} - \kappa_{Y} \lambda_{T}).
\end{eqnarray}
Due to the particular physical conditions in PNSs
($\omega_m \ll 1$), even a relatively low magnetic field can make this 
condition not satisfied. 
We must point out, however, that this criterium
is associated to the change of sign of new the magnetic mode (dashed line
in Figure 2), which is irrelevant, because the dominant unstable mode is 
always that associated to one of the other two conditions.

Let us turn back to the other two conditions, (\ref{mc2}) and (\ref{mc3}).
The dissipative frequencies are usually small compared to the 
dynamical frequencies, $\omega_{g}$ and $\omega_{L}$, 
for typical lengthscales in a PNS. For example, the 
contribution of terms proportional to the cube of dissipative 
frequencies in equations (\ref{mc2})--(\ref{mc3}) does not exceed  10\%. 
Therefore, we can neglect the 
terms containing only dissipative frequencies compared to those 
containing the dynamical ones (including the Alfv\'en frequency). This 
latter approximation is justified if the magnetic field is large enough 
for $\omega_A$ to be larger than any dissipative frequency, 
$\omega_{A} > \omega_{\rm diss} =\max (\omega_{\nu}, \omega_{T}, \omega_{Y}, 
\omega_{TY}, \omega_{YT})$. Since $\omega_{\rm diss} \sim 0.1-1$ s$^{-1}$, 
the condition $\omega_{A} > \omega_{\rm diss}$ is
fulfilled if the magnetic field exceeds some characteristic value
which is of the order of $10^{12}-10^{13}$ G.
For $B<10^{13}$ G  the influence of the magnetic field is insignificant 
and conditions (\ref{mc2})--(\ref{mc3}) are exactly equivalent to 
conditions (\ref{nfcond})--(\ref{smcond}). For $B>10^{13}$ G, the 
instability conditions can be simplified as follows:
\begin{eqnarray}
- \omega_{g}^{2} (\omega_{Y} - \alpha \omega_{YT})
- \omega_{L}^{2} \left(\omega_{T} - \frac{\omega_{TY}}{\alpha} \right)
+ \omega_{A}^{2} (\omega_{T} + \omega_{Y}) <0, 
\label{cmag1}
\\
- \omega_{g}^{2} (\omega_{T} + \omega_{\nu} + \alpha \omega_{YT})
- \omega_{L}^{2} \left(\omega_{Y} + \omega_{\nu} + 
\frac{\omega_{TY}}{\alpha} \right) 
+ \omega_{\nu} \omega_{A}^{2} < 0.
\label{cmag2}
\end{eqnarray}
As for equation (\ref{cq0}), the minimal values of the gradients which 
cause the instability correspond to $k \approx Q$ 
(short wavelengths). Substituting
$k=Q$ into equations (\ref{cmag1}) and (\ref{cmag2}), we have
\begin{eqnarray}
\frac{dY}{dz} (\delta \kappa_{T} - \beta \kappa_{Y} ) -
\frac{\Delta \nabla T}{T} (\beta \lambda_{Y} - \delta \lambda_{T})  
< - \frac{\pi B^{2}}{4 g \rho d^{2}} (\kappa_{T} + 
\lambda_{Y}),
\label{mcrit1}
\end{eqnarray}
\begin{eqnarray}
\frac{d Y}{dz} [ \delta ( \nu + \lambda_{Y} ) +  \beta \kappa_{Y}] 
-  \frac{\Delta \nabla T}{T} [ \beta ( \nu + \kappa_{T} ) + 
\delta \lambda_{T}]
< - \frac{\pi \nu B^{2}}{4 g \rho d^{2}}.
\label{mcrit2}
\end{eqnarray}
Note that these criteria are analogous to the instability criteria
for non-magnetic PNSs (\ref{nfcond})--(\ref{smcond}), but replacing
\begin{eqnarray}
\nu (\kappa_{T} \lambda_{Y} - \kappa_{Y} \lambda_{T}) \rightarrow 
\frac{d^{2} B^{2}}{27 \pi^{3} \rho} (\kappa_{T} + \lambda_{Y}), \\
(\kappa_{T} + \lambda_{Y}) [(\nu + \kappa_{T})(\nu + \lambda_{Y}) -
\lambda_{T} \kappa_{Y}] \rightarrow \frac{d^{2} B^{2}}{27 \pi^{3} \rho} 
\nu, 
\end{eqnarray}
and the magnetic field plays the role of an efficient dissipative
process. In magnetic PNSs the dissipative effect caused by the Lorentz force 
is much greater than neutrino transport or neutrino viscosity effects.

In Figure 3 we show the unstable zones at each time of the evolution
of a young PNS from one of the numerical simulations in
\cite{Pons99}. Top and bottom panels correspond to criteria 
(\ref{mcrit1}) and (\ref{mcrit2}), respectively. 
Criterion (\ref{mcrit2}) depends on the neutrino viscosity, for which
available estimates are sometimes contradictory 
\citep{vH81,GP82,TD93}.
In our simulations we used the value $\nu = \eta^{2} \kappa_{T}$ 
\citep{vH81}.
The contours indicate the convectively unstable region for
the following values of a passive magnetic field: $10^{14}$ G (solid), 
$5\times 10^{15}$ G (dotted), 
$10^{16}$ G (dashed), $3\times 10^{16}$ G (dash-dot) and $5\times 10^{16}$ 
(dash-dot-dot-dot).
It turns out that, with this viscosity, 
the critical magnetic fields for criteria (\ref{mcrit1}) and (\ref{mcrit2}) 
are of the same order of magnitude. The magnetic correction
on (\ref{mcrit2}) is somewhat more restrictive and, 
generally, its associated instability is suppressed with
weaker magnetic fields. The difference is caused mainly by the 
different combinations of kinetic coefficients that appear in the
equations.
Both criteria require a rather strong magnetic 
field, $B \sim (1-5) \times 10^{16}$G, in order to stabilize
the fluid completely, although
the region where condition (\ref{mcrit2}) is fulfilled is typically
smaller than that given by condition (\ref{mcrit1}). Note that 
moderate magnetic fields in the order of $~\leq 10^{14}$ G leave
the unstable regions essentially unchanged, and the effects are visible
only for  $B \geq 10^{15}$ G.

In Figure 4 we plot the critical magnetic field 
($B_{\rm st}$) which completely stabilizes a certain layer
of the PNS (corresponding to enclosed baryonic masses of 
0.2, 0.5, 1.0, and 1.4 $M_{\odot}$) as a function of time.
Layers with $M_{\rm B} < 1M_{\odot}$ are stable at the beginning of the
evolution.  Later on, when the corresponding 
layer becomes unstable, $B_{\rm st}$ grows rapidly, reaching some
maximum value, and then decreases monotonously as the 
temperature and lepton gradients are smeared out via neutrino diffusion.  
The maximum value of $B_{\rm st}$ at different points varies in a very narrow 
range ($2-5 \times 10^{16}$ G), but 
this maximum is reached at different instants in time. Note that 
$B_{\rm st}$ provides stabilization in the largest unstable lengthscale.
Stabilization at smaller scales is accomplished with weaker magnetic 
fields. The instability seems to be especially persistent in the 
central region with $M_{\rm B} < 0.5 M_{\odot}$, where convective motions
might last for a long time \citep{KJM96} and 
$B_{\rm st} \sim 5 \times 10^{16}$.

\section{Discussion}

In this paper we have considered the effect of magnetic field
(consistently coupled to other dissipative processes, such as neutrino 
transport and neutrino viscosity) on the convective instability of PNSs. 
Our results show that transport phenomena play an important role in
determining the size of the region of the star which is subject to 
instability. The particular values of the transport and kinetic 
coefficients also determine the kind of instabilities
which can be present in an unstable region ({\it convection, neutron
fingers, semiconvection}). 

For low magnetic fields ($B<10^{13}$ G)
the region where instabilities are present remains unaffected when
compared to the non-magnetic case. That region
covers an important part of the PNS and lasts for almost the whole 
Kelvin-Helmholtz phase of the PNS evolution. Consequently,
due to the existence of such an extended region subject to instabilities,
convective motions are likely to appear and
turbulent dynamo action can significantly increase, locally, 
the magnetic field strength.

Once the magnetic field has a significant strength ($10^{13}$ G 
$<B<10^{15}$ G), its effect is twofold. First, it is
responsible for the appearance of a new mode which can generally
be unstable, although in this case, its growth time is always longer 
than that of the other unstable modes.
Secondly, the magnetic field acts as a stabilizing 
agent for all types of instability studied in the non-magnetic case,
progressively reducing the size of the convective regions.
Our calculations show, however, that only a relatively strong magnetic field
can influence significantly the stability properties of PNSs, 
particularly in the central region with $M_{\rm B} < 0.5 M_{\odot}$. 
For instance, if the dynamo action is sufficiently efficient, 
the duration of the convective epoch can be noticeably 
shorter as the magnetic field increases up to $\geq 10^{16}$G. 
We find that the critical magnetic field that suppresses all sorts
of instability happens to be in a narrow range, at about
$5 \times 10^{16}$ G. This can be considered as an upper limit
to the poloidal magnetic field generated by convective dynamo.

In our simplified approach, we have considered the case of a 
vertical magnetic field.  It is known \citep{Cha61} that this magnetic 
configuration requires the maximal field strength for stabilization.
In real conditions, the field geometry may be more complex with a 
substantial component perpendicular to the temperature and composition 
gradients. In PNSs, this field component can correspond, 
for example, to a toroidal magnetic field. For this geometry, the 
stabilizing magnetic field is typically several times lower. 
Therefore, stabilization against hydromagnetic instabilities can 
really take place at a slightly weaker magnetic field than that 
obtained in our calculations.  In the presence of differential 
rotation (a very likely situation after the core collapse), the main 
contribution to the total field strength would be given 
by the toroidal magnetic field ($B_{\rm t}$) and the poloidal
component ($B_{\rm p}$) would be less important. In a quasi-stationary 
regime, the rate of stretching of the toroidal field lines from the 
poloidal configuration by differential rotation is compensated by 
turbulent dissipation \citep{Par79}.
By equating both rates one can write the following equation: 
\begin{equation}
\frac{B_{\rm p}}{\tau} \sim \frac{v_{T} \ell_{T}}{3} \nabla^2 B_{\rm t},
\end{equation}  
where $\tau$ is the characteristic stretching time,
$v_{T}$ the turbulent velocity and $\ell_{T}$ the turbulent lengthscale. 
Since the stretching is due to differential
rotation, $\tau$ might be estimated as $\tau = \mid\Delta\Omega\mid^{-1}$,
$\Delta \Omega$ being the difference between the maximum and minimum angular 
velocity. An estimation of $\nabla^2 B_{\rm t}$ is $\sim B_{\rm t}/R^2$, 
$R$ being the typical radius of the star. Using the following values, 
$R \sim 20$ km, $v_{T} \sim 10^{8}$ cm/s and $\ell_{T} \sim 
10^{5}$cm, we obtain 
\begin{equation}
\frac{B_{t}}{B_{p}} \sim \frac{\Delta \Omega}{1  {\rm s}^{-1}}.
\end{equation} 
Assuming that neutron stars are born with large angular velocities
(say $\Omega = 10^{3}$ s$^{-1}$), and that $\Delta \Omega / \Omega$
is comparable to that of the Sun ($0.1-0.3$),
we can estimate the toroidal field to be $\approx 100-300$ times
stronger than the poloidal component.  Since the total magnitude of 
the magnetic field generated by dynamo action should not exceed, 
in any case, $10^{16}$ G we have to conclude that
the poloidal component which may represent the observed pulsar field
must be in the range $3 \times 10^{13} - 10^{14}$ G. This estimate is
in agreement with the observed fields of young pulsars
in supernova remnants. The toroidal field could be, however, 
much stronger, which might lead to important consequences on the structure 
of neutron stars \citep{Boc95,CPL01}.

\acknowledgements

This work was supported by 
the Spanish Ministerio de Ciencia y Tecnolog\'{\i}a (grant AYA 2001-3490-
C02-02), the EU Program `Improving the Human Research Potential and 
Socio-Economic Knowledge Base' (Research Training Network contract 
HPRN-CT-2000-00137), the Russian Foundation
of Basic Research and Deutsche Forschungsgemeinschaft (grant 
00-02-04011). V.U. also thanks the Ministerio de Educacion, Cultura
y Deporte of Spain for financial support under the grant SAB1999-0222.
J.P. is supported by the Marie Curie Fellowship No. HPMF-CT-2001-01217.
The authors acknowledge to an anonymous referee for useful suggestions.

\clearpage

\newpage

\begin{table}  
\begin{center}  
\caption{Physical conditions and transport coefficients for three
representative situations during the first seconds of life of a PNS.
Cases A,B,C represent the three possibilities found in numerical 
simulations, classified according to criteria (\ref{nfcond})--(\ref{smcond}). }
\vspace*{0.15in}  
\begin{tabular}{|l|ccccc|cccc|}  
\hline  
\multicolumn{1}{|c|}{Case} & \multicolumn{5}{c|}{Thermodynamic state}  
& \multicolumn{4}{c|}{Transport coefficients (cm)} \cr  
\hline  
{} & $n_B$ (fm$^{-3}$) & T (MeV) & $Y_L$ & $\beta$ & $\delta$  
& $\kappa_T$ & $\kappa_Y$ & $\lambda_T$ & $\lambda_Y$ \cr  
\hline  
A (stable) & 0.301 & 21.1 & 0.333 & 0.103 & 0.111  & 0.556 & -0.851 & -0.046 & 0.496  \\  
B (convection) & 0.350 & 33.6 & 0.259 & 0.145 & -0.058 & 0.436 & -0.536 & -0.038 & 0.285 \\  
C (fingers) & 0.197 & 12.1 & 0.096 & 0.069 & -0.570 & 2.46 & -20.1 & -0.23 & 4.67 \\  
\hline  
\end{tabular}  
\end{center}  
\end{table}  

\clearpage
\newpage


\figcaption{\label{figure1}}
{Stability plane for the three thermodynamic states detailed
in Table 1. Four different regions in the plane are separated by
criteria (\ref{nfcond})-(\ref{smcond}) (solid and dotted lines, 
respectively). The shaded area indicates the region where the three
roots of equation (\ref{cubic}) are real.
The particular values of the gradients in each studied case (A,B,C) 
are indicated by asterisks.  The scale factor
$\ell$ has been taken as the pressure height of each case.
}

\figcaption{\label{figure2}}
{Sketch showing the behaviour of the real part of the 
roots of equation (\ref{quartic}) as a function of the square of 
the Alfv\'en frequency, for fixed values of the dynamical and dissipative
frequencies. Dashed line corresponds to the mode decoupled in the
non-magnetic case (referred to as {\it magnetic mode} in the text). 
Units are arbitrary and the points $a$, $b$ and $c$ indicate changes
of sign and character (real--complex) of the modes.
See \S 3.1 for further details.
}

\figcaption{\label{figure3}}
{Contour plots showing the variation of the different convectively 
unstable zones, assuming different strengths of the magnetic field.
The contour lines correspond to the following values of the
magnetic field: $10^{14}$ G (solid), $5\times 10^{15}$ G (dotted), 
$10^{16}$ G (dashed), $3\times 10^{16}$ G (dash-dot) and $5\times 10^{16}$ 
(dash-dot-dot-dot).
The top panel refers to the unstable region according to
criteria (\ref{mcrit1}) and the bottom panel refers to (\ref{mcrit2}).
The x-axis indicates the evolution time and the y-axis the enclosed
baryonic mass of the layer. The innermost part of the star 
remains unstable at intermediate times even when magnetic fields 
as large as 10$^{16}$ G are present.
The asterisks correspond to the cases A, B and C described in the text.
This PNS model was taken from numerical simulations performed in
\cite{Pons99}.
}

\figcaption{\label{figure4}}
{Critical magnetic field strength that completely stabilizes
the fluid as a function of the evolution time. The four
lines stand for four different layers, with enclosed baryonic
masses of 0.2 (solid), 0.5 (dotted), 1.0 (dashed) and 1.4 (dash-dot)
$M_\odot$, respectively. The PNS model is the same as in Figure 3.
In general, the deeper is the layer, the larger magnetic field
is required for unconditional stabilization.
}

\newpage

\begin{figure}[t]
\epsfxsize=10cm \epsfbox{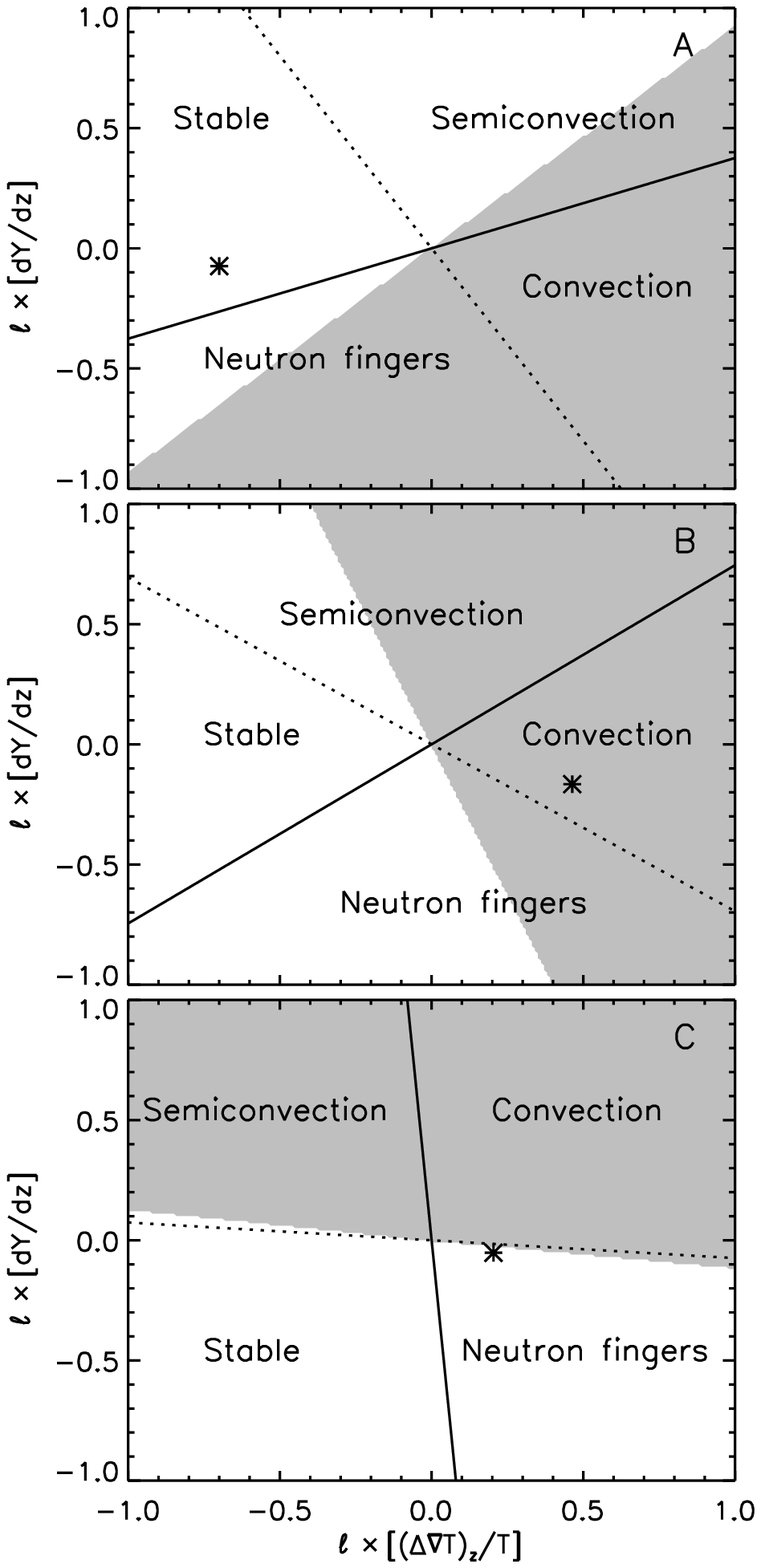}
\end{figure}
 
\newpage
  
\begin{figure}[t]
\epsfxsize=14cm
\epsfysize=18cm
\epsfbox{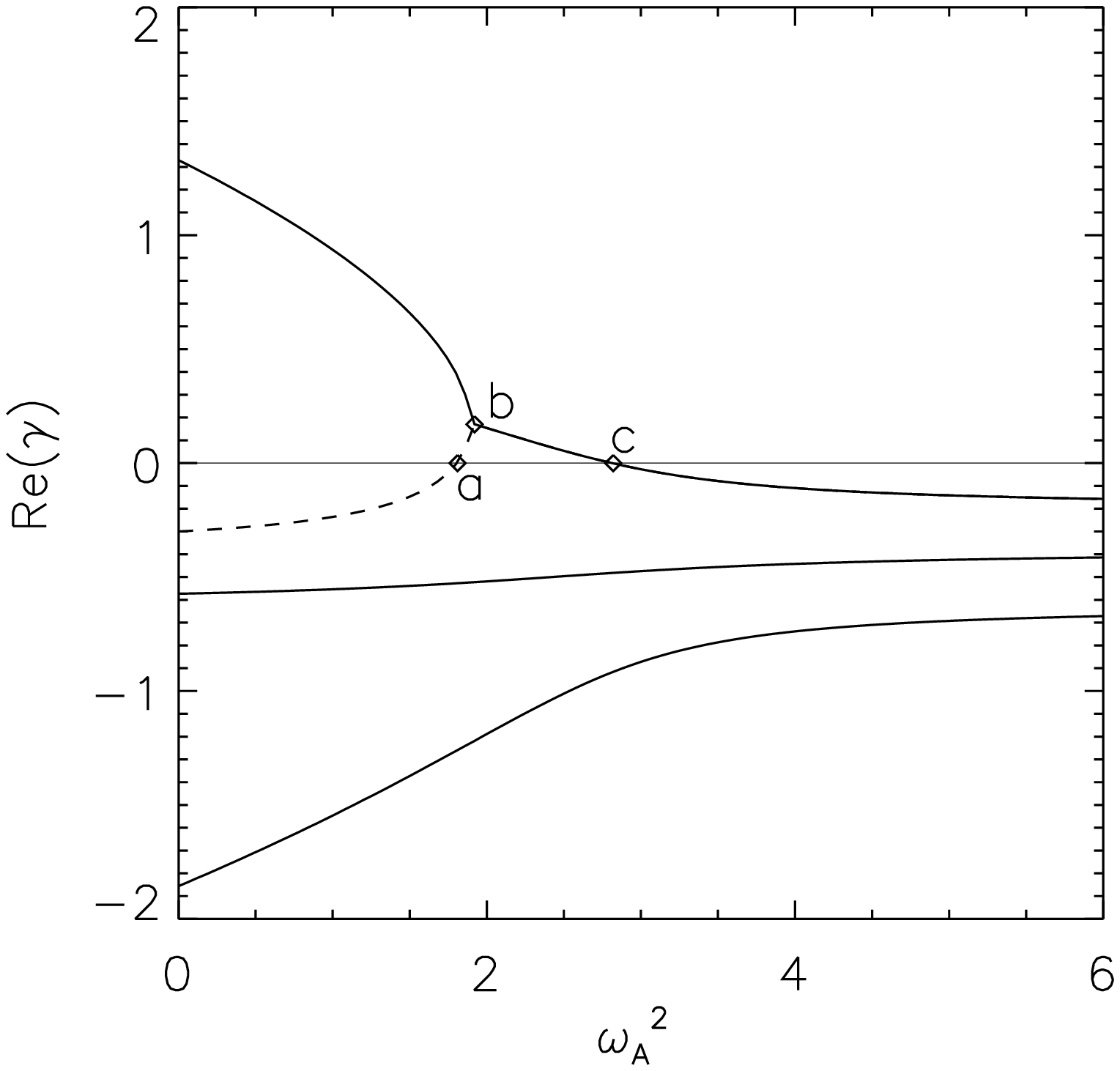}
\end{figure}
   
\newpage
    
\begin{figure}[t]
\epsfbox{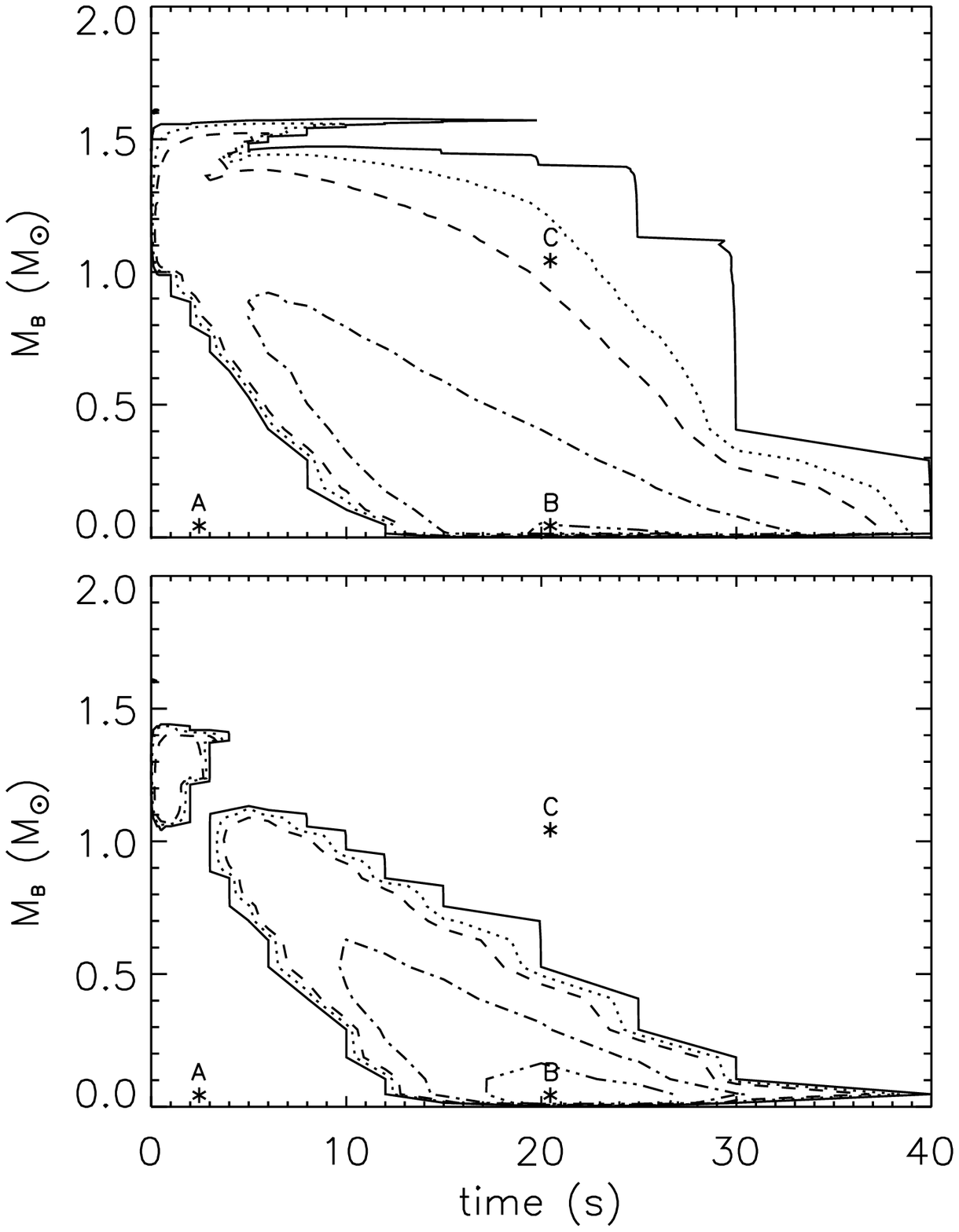}
\end{figure}
     
\newpage
      
\begin{figure}[t]
\epsfbox{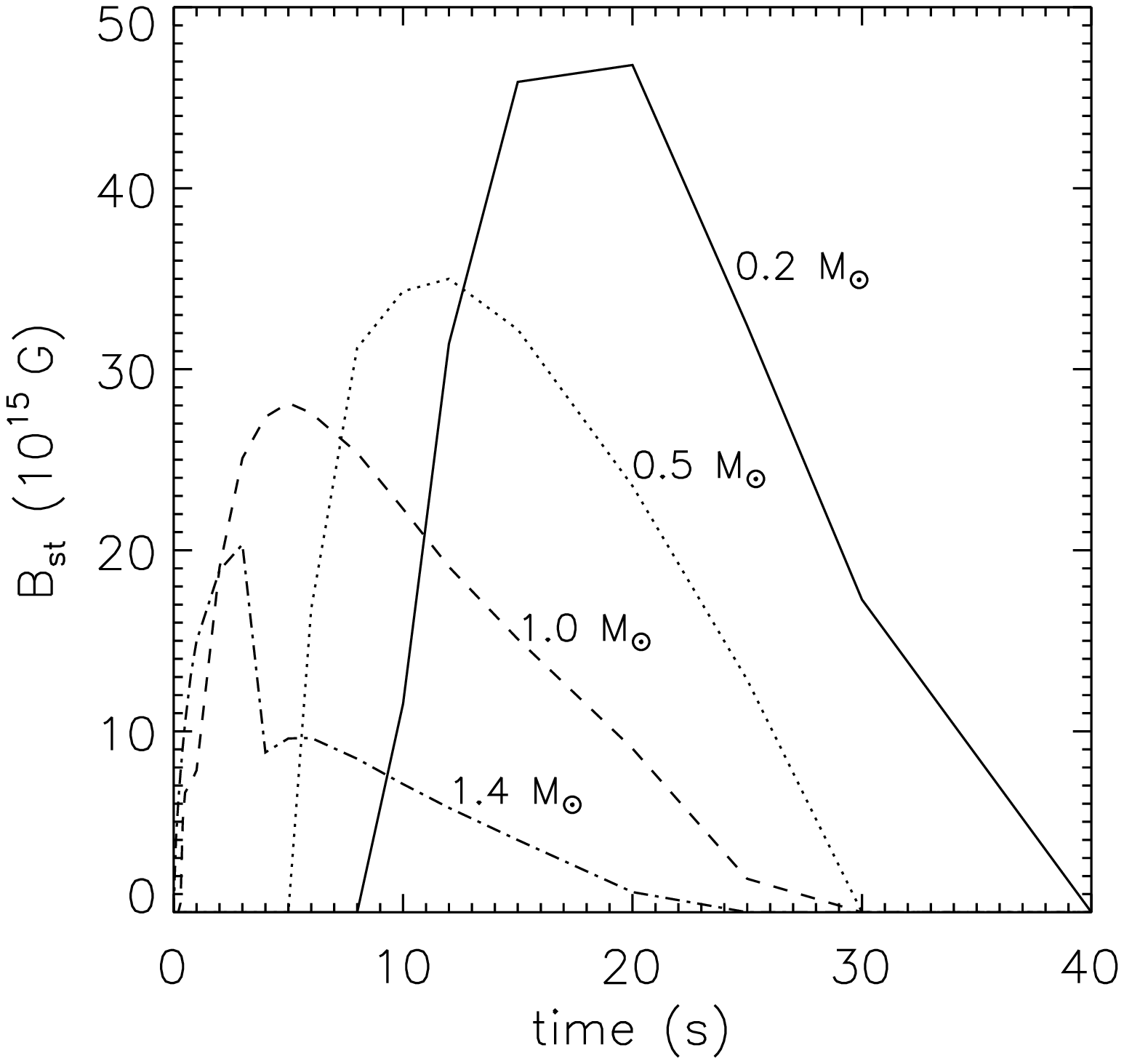}
\end{figure}

\end{document}